\documentclass[a4paper,aps,prl,twocolumn,a4paper]{revtex4}
\usepackage{epsf}
\usepackage{graphicx}
\usepackage{bm}
\usepackage[T1]{fontenc}
\usepackage{amsmath,amssymb}
\usepackage[latin1]{inputenc}
\usepackage{times}
\usepackage{ulem}
\usepackage{color}
\usepackage[squaren,Gray]{SIunits}
\newcommand{\cora}[1]{\textcolor{black}{#1}}

\begin{document}

\title{Retrieving time-dependent Green's functions in optics with low-coherence interferometry}

\author{Amaury Badon}
\author{Geoffroy Lerosey}
\author{Albert C. Boccara}
\author{Mathias Fink}
\author{Alexandre Aubry}
\affiliation{Institut Langevin, ESPCI ParisTech, PSL Research University, CNRS UMR 7587, 1 rue Jussieu, 75005 Paris, France}
\email{alexandre.aubry@espci.fr}

\date{\today}

\begin{abstract}
We report on the passive measurement of time-dependent Green's functions {in the optical frequency domain} with low-coherence interferometry. Inspired by previous studies in acoustics and seismology, we show how the correlations of a broadband and incoherent wave-field can directly yield the Green's functions between scatterers of a complex medium. Both the ballistic and multiple scattering components of the Green's function are retrieved. This approach opens important perspectives for optical imaging and characterization in complex scattering media.
 \end{abstract}


\maketitle

Waves propagating in complex media can experience complicated trajectories through scattering off objects or reflection and refraction at interfaces. All these events are nonetheless captured by the Green's function formalism. \cora{Mathematically, the Green's function is the solution of the wave equation with a point source term \cite{nieto}. It connects the wave-field to any excitation by means of time and space convolutions with the source distribution.} In random multiple scattering and reverberating media, the temporal Green's function thus provides a unique signature of the propagation of waves between the source and observation points. This property has been put to profit to focus waves in acoustics and electromagnetism, where temporal Green's function are easily accessible, through the concept of time reversal \cite{derode0,lerosey}. Similarly, temporal Green's functions have been proved to allow imaging of complex media, either using statistical approaches or through numerical reconstruction \cite{borcea,aubry2009random, larose2, krishnan}. In optics, there have been recently exciting proposals to measure Green's functions of complex media \cite{popoff2010measuring,choi,popoff2,choi2}, and use them for imaging or focusing purposes, mostly owing to the development of wave front shaping techniques \cite{vellekoop2007focusing,Mosk_review}. 

Incidentally, previous studies in acoustics have proposed a simpler and elegant route towards a passive measurement of temporal Green's functions without the use of any source \cite{weaver,derode}.
%
%
The cross-correlation (or mutual coherence function) of an incoherent wave-field measured at two points A and B can yield the time dependent Green's function between these two points. Provided that the ambient field is equipartitioned in energy, the time derivative of the correlation function at two positions is actually proportional to the difference between the anticausal and causal Green's functions.
This property is a signature of the universal fluctuation dissipation theorem \cite{rytov,agarwal,bart} and has been derived following different approches \cite{weaver,bart,wapenaar,snieder,wapenaar2,larose}. An elegant physical picture is provided by an analogy with time reversal \cite{derode,derode2,supp}.  In the frequency domain, this result manifests itself as the link between the spatial correlation of the wave-field and the local density of states \cite{Joulain,caze}. This fundamental quantity is actually proportional to the imaginary part of the monochromatic \textit{self}-Green's function.

Previously and independently developed in helioseismology \cite{duvall}, the Green's function estimation from diffuse noise cross correlations has received a considerable attention in seismology in the 2000s \cite{campillo}. The cross-correlation of seismic noise recorded by two stations over months has allowed to retrieve the Green's functions between these observation points as if one was replaced by a virtual coherent source. By passively measuring the elastic Green's functions between a network of seismic stations, an imaging of the Earth's crust has been obtained with unprecedented high resolution \cite{shapiro}. More recently, thermal radiation noise has also been taken advantage of to measure passively electromagnetic Green's functions in the microwave {frequency domain} \cite{davy}.

The aim of this paper is to demonstrate the passive measurement of time-dependent Green's functions in optics. For this proof-of-concept, the first sample under study consists in dispersed microbeads that are used as passive sensors. This scattering sample is isotropically illuminated by an incoherent halogen light source. The correlation of the scattered wave-field is measured by means of a Michelson interferometer and recorded on a CCD camera. In this Letter, we first show that the cross-correlation of the wave-field coming from two scatterers $A$ and $B$ converges towards the ballistic Green's function $g_{AB}(t)$ between them. In a second example, we show that the multiple scattering components of the Green's function can also be retrieved. This is illustrated by measuring the autocorrelation of the field coming from the scatterer $A$. The resulting \textit{self}-Green's function $g_{AA}(t)$ is shown to exhibit the time-resolved multiple scattering echoes between the scatterers $A$ and $B$. At last, we show that our approach can also be extended to a strongly scattering medium made of ZnO {nanoparticles}. The correlation of the scattered wave-field directly yields the time-dependent Green's functions between each pixel of the CCD camera. With a moderate integration time, the resulting Green's functions are shown to emerge from noise for times of flight at least twenty five times larger than the transport mean free time. Hence, this simple and powerful approach directly yields a wealth of information about the scattering medium. It opens important perspectives for imaging and characterization in strongly scattering media.
%
%

\begin{figure}[htbp]
\includegraphics[width=8.5cm]{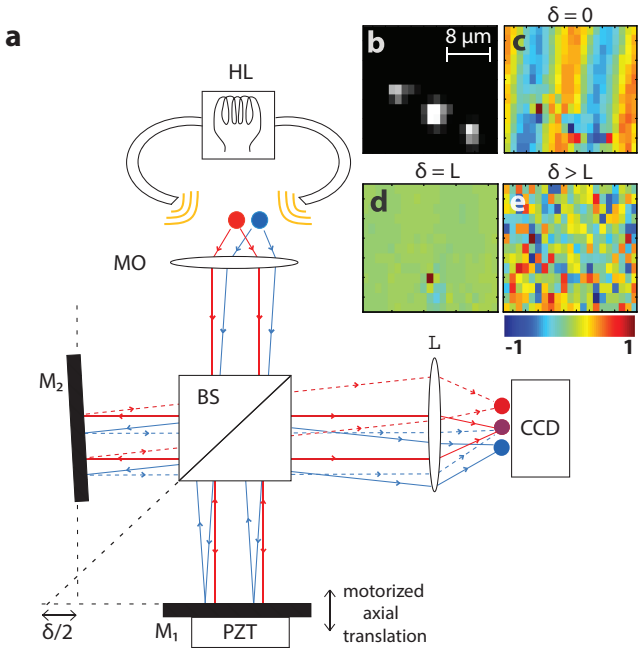}
\caption{(a) Experimental set up. A broadband incoherent light source isotropically illuminates a scattering sample (here consisting of two beads). The spatio-temporal correlation of the scattered wave field is extracted by means of a Michelson interferometer and recorded by a CCD camera. HL : halogen lamp. MO : microscope objective. BS : beam splitter. M : mirror. PZT : piezoelectric actuator. (b) Intensity image recorded by the CCD camera displaying the two beads and their superimposed image at the center. (c,d,e) Field measured by the CCD camera \textit{via} phase-shifting interferometry at different optical path difference $\delta$: (c) $\delta=0$, (d) $\delta=L$, (e) $\delta>L$. Each field has been normalized by its maximum {range}.}
\label{fig1}
\end{figure}
The experimental set-up is displayed in Fig.\ref{fig1}. An incoherent broadband light source (650-850 nm) isotropically illuminates a scattering sample {in a \textit{dark-field} configuration}. \cora{The coherence properties of the incident wave-field are shown in the Supplemental Material \cite{supp}. The incident wave-field exhibits a coherence time {$\tau_c \sim 10 $ fs} and a coherence length {$l_c \sim 3$ $\mu$m} }. The scattered wave-field is collected by the microscope objective and sent to a Michelson interferometer. The beams coming from the two interference arms are then recombined and focused by a lens. The intensity recorded by the CCD camera at the focal plane can be expressed as,
\begin{equation}
S(\mathbf{r},\mathbf{r}+\Delta \mathbf{r},t) =  \int_{0}^T || \mathbf{E}(\mathbf{r},t+\tau)+\mathbf{E}(\mathbf{r}+\Delta \mathbf{r},\tau) ||^2  \mathrm{d}\tau 
\label{intensity}
\end{equation}
with $\tau$ the absolute time, $\mathbf{r}$ the position vector on the CCD screen, $\mathbf{E}(\mathbf{r},\tau)$ the scattered electric field associated to the first interference arm and $T$ the integration time of the CCD camera. The tilt of mirror $M_2$ allows a displacement $\Delta \mathbf{r}$ of the associated wave-field on the CCD camera. The motorized translation of mirror $M_1$ induces a time delay $t = \delta/c$ between the two interferometer arms, with $\delta$ the optical path difference (OPD) and $c$ the light celerity. The interference term is extracted from the intensity pattern (Eq.\ref{intensity}) by phase shifting interferometry (``four phases method'' \cite{popoff2010measuring}) using a piezoelectric actuator placed on mirror $M_1$. It directly yields the mutual coherence function $C$ of the scattered wave-field $\mathbf{E}$:
\begin{equation}
C(\mathbf{r},\mathbf{r}+\Delta \mathbf{r},t) = \int_{0}^T \mathbf{E}(\mathbf{r},t+\tau) \cdot \mathbf{E}(\mathbf{r}+\Delta \mathbf{r},\tau)  \mathrm{d} \tau
\label{cor}
\end{equation}
If the incident light is spatially and temporally incoherent, the time derivative of the correlation function $C(\mathbf{r_A},\mathbf{r_B},t)$ between two points A and B should converge towards the difference between the causal and anti-causal Green's function, such that
\begin{equation}
\partial_t C(\mathbf{r_A},\mathbf{r_B},t) \underset{T\rightarrow \infty}{\sim}  g_{AB}(t) - g_{AB}(-t)  .
\label{green}
\end{equation} 
\cora{A theoretical proof for this fundamental result is provided in the Supplemental Material \cite{supp}. Our argument is based on symmetries of reciprocity, time-reversal invariance and diffraction theory.}

The aim of this Letter is to prove experimentally this result and measure time-dependent Green's functions with the basic experimental setup displayed in Fig.\ref{fig1}. As a proof-of-concept, we first study a sample made of 3$\mu$m-diameter magnetite beads (Fe$_3$O$_4$) randomly embedded in a transparent polymer matrix (poly-L-lysine) on a microscope slide. The correlation function between two scatterers can be measured by tilting mirror $M_2$ such that the images of the two beads are superimposed on the same CCD camera pixel [see Fig.\ref{fig1}(b)]. Figs.\ref{fig1}(c,d,e) display the interference pattern recorded by the CCD camera at different OPD for an isolated couple of beads, $A$ and $B$, separated by a distance $L=$7 $\mu$m. At $\delta=0$, the straight fringes observed on the CCD camera result from a residual coherence of the incident wave-field [Fig.\ref{fig1}(c)]. At $\delta=L$, a strong interference signal is observed in the area where the images of the two beads overlap [Fig.\ref{fig1}(d)]. As we will see, it corresponds to the direct echo between $A$ and $B$. For $\delta>L$, the field measured by the CCD camera corresponds to noise that results from the interference between uncorrelated random wave-fields [Fig.\ref{fig1}(e)]. 
\begin{figure}[htbp]
\centering
\includegraphics[width=8.5cm]{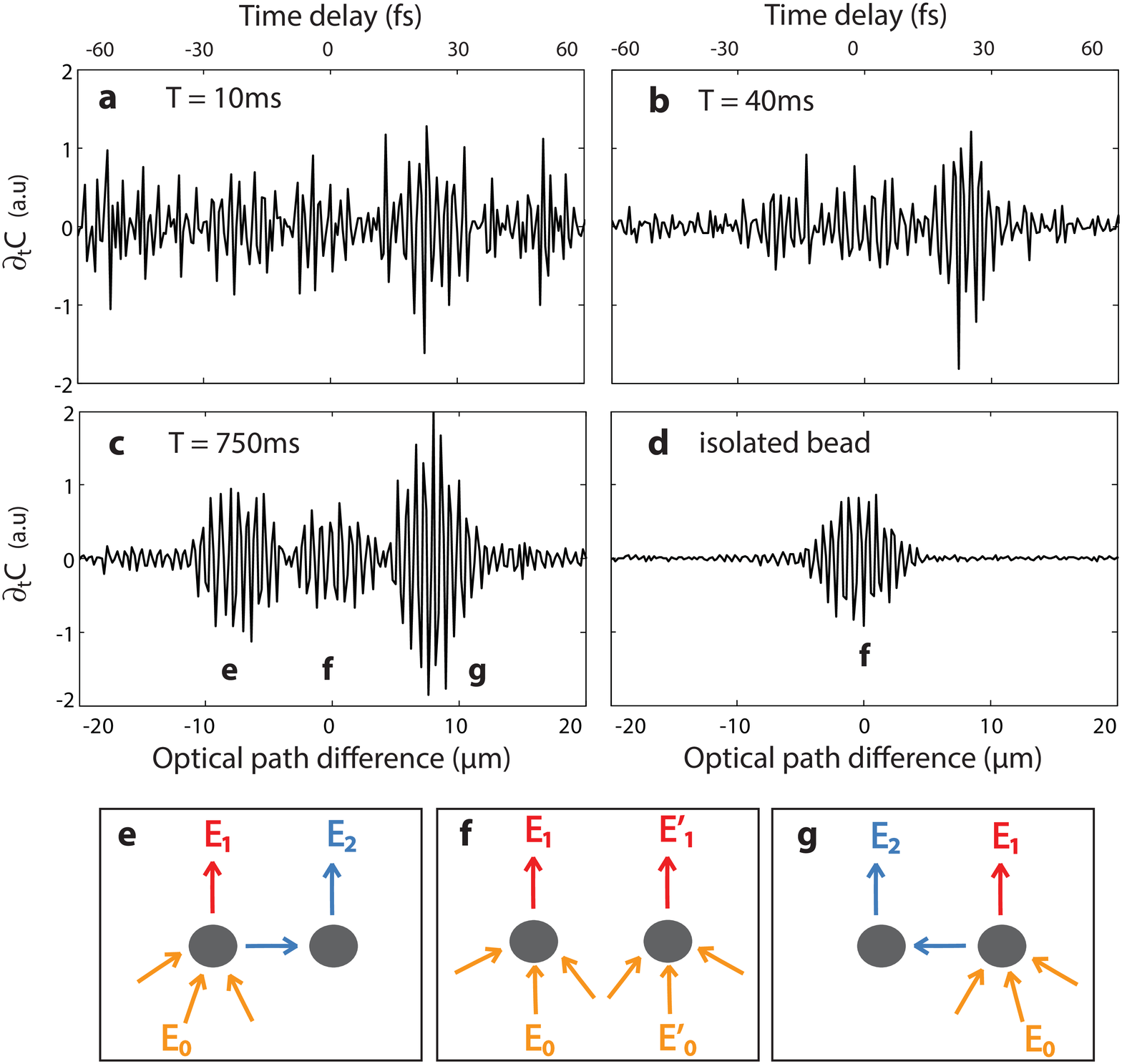}
\caption{Passive measurement of the Green's function $g_{AB}(t)$ between the two beads $A$ and $B$. (a,b,c) Cross-correlation signals versus time delay / OPD for different integration times: $T=$10 ms (a), $T=$40 ms (b) and $T=$750 ms (c). (d) Interferometric signal obtained for the isolated bead $D$ ($T=750$ ms). (e,f,g) Sketch of the scattering events accounting for the different pulses emerging from the signals in (c) and (d). }
\label{fig2}
\end{figure}

Fig.\ref{fig2} displays the time-dependence of the cross-correlation function between beads $A$ and $B$ for different integration times. This interferometric signal contains a random contribution that should vanish with average and a deterministic contribution due to the stationary interferences between the two beads. The latter one should directly lead to the Green's function $g_{AB}(t)$ as stated by Eq.\ref{green}. For $T=10$ ms, noise predominates and no coherent signal can be clearly detected [Fig.\ref{fig2}(a)]. For $T=40$ ms, three time-resolved echoes start to emerge from noise but the signal-to-noise ratio is still perfectible [Fig.\ref{fig2}(b)]. At last, for $T=750$ ms, noise is sufficiently averaged out to obtain the stationary interference signal with a good precision [Fig.\ref{fig2}(c)]. As a reference, Fig.\ref{fig2}(d) displays the cross-correlation function obtained between an isolated bead $D$ and the background wave-field for the same integration time and tilt of $M_2$. 
Each interferometric signal in Figs.\ref{fig2}(c,d) exhibits an echo around $\delta=0$ that corresponds to the straight fringes displayed by Fig.\ref{fig1}(c). Although the two beads $A$ and $B$ are separated by a distance $L>l_c$, the incident wave fields, $E_0$ and $E'_0$, seen by each of them remain slightly correlated \cora{\cite{supp}}. This gives rise to a stationary interference between the single scattering paths $E_1$ and $E'_1$ around $\delta=0$ [Fig.\ref{fig2}(f)]. Unlike the signal associated to the isolated bead $D$ [Fig.\ref{fig2}(d)], the cross-correlation between beads $A$ and $B$ clearly exhibits two echoes around $\delta$ = $\pm L$ [Fig.\ref{fig2}(c)]. It corresponds to the strong interference signal previously highlighted in Fig.\ref{fig1}(d). These two echoes are the expected causal and anticausal parts of the ballistic Green's function between $A$ and $B$. They originate from the interference between the single scattering path $E_1$ and the double scattering path $E_2$ depicted in Figs.\ref{fig2}(e,g). Normally, the causal and anticausal parts of the Green's function should be of same amplitude due to reciprocity. \cora{This is not strictly the case here because the illumination is not perfectly isotropic.}

\begin{figure}[htbp]
\centering
\includegraphics[width=8.5cm]{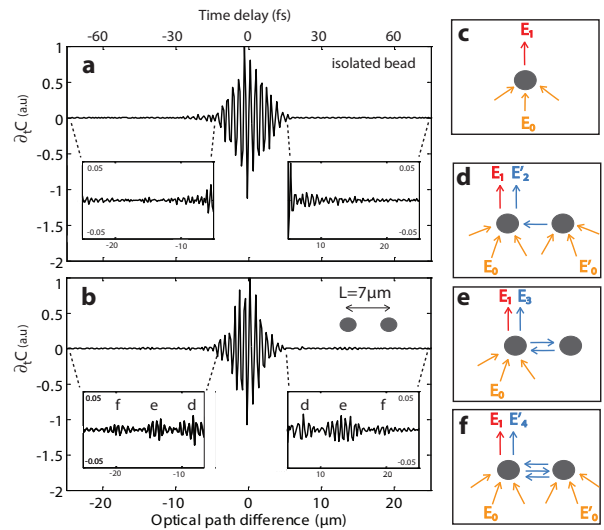}
\caption{Passive measurement of the \textit{self}-Green's function $g_{AA}(t)$ associated to the bead $A$. (a)-(b) Autocorrelation signal versus time delay / OPD for the isolated bead $D$ (a) and the bead $A$ placed at the vicinity of bead $B$ (b). (c)-(f) Sketch of the scattering events accounting for the different pulses emerging from the autocorrelation signals in (a) and (b).}
\label{fig4}
\end{figure}
This first experiment has demonstrated the ability of measuring a direct echo between two scatterers with low-coherence interferometry. However, one can go beyond and also measure the multiple scattering components of the Green's function. As a demonstration, we show the measurement of the autocorrelation function associated to the bead $A$. This is performed by simply canceling the tilt of mirror $M_2$ such that $\Delta \mathbf{r}$ = 0 in Eq.\ref{cor}. The autocorrelation function $C(\mathbf{r_A},\mathbf{r_A},t)$ should lead to a measurement of the \textit{self}-Green's function $g_{AA}(t)$ [Eq.\ref{green}]. In this case, the bead $A$ virtually acts both as the source and the receiver while the bead $B$ only acts as a passive scatterer. Our aim is to detect the presence of bead $B$ in $g_{AA}(t)$ through the multiple scattering echoes that take place between the two beads. Fig.\ref{fig4} compares the autocorrelation signal obtained for the bead $A$ [Fig.\ref{fig4}(b)] and for the isolated bead $D$ [Fig.\ref{fig4}(a)]. In the latter case, the autocorrelation function only exhibits a single echo around $\delta=0$ due to the interference of the scattered field $E_1$ with itself [Fig.\ref{fig4}(c)]. For bead $A$, six supplementary echoes are visible at $\delta$ = $\pm L$, $\delta$ = $\pm2L$ and $\delta$ = $\pm3L$. Each of them can be associated to a stationary interference between multiple scattering paths depicted in Figs.\ref{fig4}(d,e,f). The signal at $\delta=\pm L$ results from the interference between the single scattering path $E_1$ and the double scattering path $E'_2$ [Fig.\ref{fig4}(d)]. It corresponds to the direct echo between the two beads. It would not emerge if the incident wave fields seen by the two beads, $E_0$ and $E'_0$, were totally uncorrelated. However, \cora{as shown in the Supplemental Material \cite{supp}}, a residual coherence subsists for the incident wave-field at such a distance $L$. The echo at $\delta=\pm 2L$ results from the interference between the single scattering path $E_1$ and the triple scattering path $E_3$ [Fig.\ref{fig4}(e)]. As these two paths involve the same first scattering event, it would exist even for a perfectly incoherent incident wave-field. This signal corresponds to a way and return echo between the two beads. At last, the echo at $\delta=\pm 3L$ results from the interference between the single scattering path $E_1$ and the quadruple scattering path $E'_4$ [Fig.\ref{fig4}(f)]. As for the echo at $\delta=L$, it is only visible because of the residual coherence of the incident wave-field. Hence, the relevant echoes here are the ones associated to a roundtrip scattering event between the beads as they would exist even for a perfectly incoherent wave-field. They correspond to the causal and anti-causal parts of the Green's function $g_{AA}(t)$ associated to bead $A$ and yield an information about its local environment, \textit{i.e} the presence of bead $B$ at a distance $L$. Hence this experiment demonstrates the ability of measuring passively the multiple scattering contribution of a temporal Green's function from an incoherent wave-field. In this example, the single, double and triple scattering events are nicely retrieved. Note that, in principle, it would be possible to measure higher order scattering events. To that aim, the integration time should be increased in order to lower the noise level but it would be at the cost of a longer measurement.

\begin{figure}[htbp]
\centering
\includegraphics[width=8.5cm]{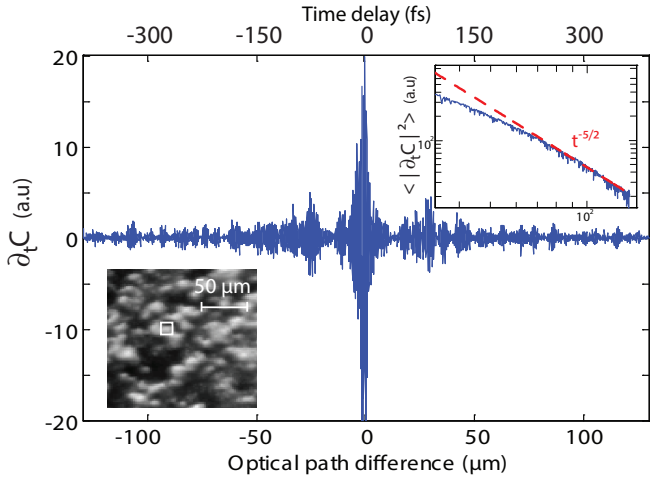}
\caption{Passive measurement of the \textit{self}-Green's function in a multiple scattering sample made of ZnO nanoparticles. The autocorrelation signal obtained for one pixel of the CCD camera is shown as a function of the time delay / OPD. The CCD image of the sample as well as the location of the selected pixel are displayed in the bottom left inset. The intensity of the autocorrelation signal averaged over 24 neighbour pixels versus time delay / OPD is shown in the top right inset in a log-log scale. The averaged intensity (continuous blue line) is fitted with the expected power law $t^{-5/2}$ (dashed red line).}
\label{fig5}
\end{figure}
Now that the ability of measuring time-dependent Green's functions between individual scatterers has been demonstrated, the case of a strongly scattering medium is now investigated. The sample under study is {a layer of slighly pressed ZnO} nanoparticles (Sigma Aldrich 544906) on a microscope slide. The sample thickness and the transport mean free path $l^*$ are of the order of 50 $\mu$m and 5 $\mu$m, respectively. The measurement of the autocorrelation signal for one pixel of the CCD camera is shown in Fig.\ref{fig5}. It directly yields the \textit{self}-Green's function $g_{AA}(t)$ for a virtual sensor $A$ placed at the surface of the sample. Its characteristic size is given by the coherence length $l_c$ of the \cora{incident} wave-field. The integration time has been fixed to 1250 ms to get rid of noise in the interference signal. The Green's function shown in Fig.\ref{fig5} is characteristic of a strongly scattering sample with a long tail that results from the numerous multiple scattering events that take place within the scattering medium. Here, we have access to a satisfying estimation of the Green's function over 130 $\mu$m in terms of OPD (or 430 fs in terms of time delay), which corresponds to scattering paths of $26 l^*$. In the multiple scattering regime, a probabilistic approach is generally adopted to extract information from the measured Green's functions. One can for instance study the intensity of the \textit{self}-Green's function averaged over several pixels of the CCD camera [see inset of Fig.\ref{fig5}]. This leads to an estimation of the \textit{return probability}, a key quantity in multiple scattering theory \cite{skipetrov} that describes the probability for a wave to come back close to its starting point. For a source placed at the surface of a scattering medium, the return probability is supposed to decrease as $t^{-5/2}$ in the diffusive regime \cite{dogariu,aubry_2014}. As shown in the top right inset of Fig.\ref{fig5}, such a power law decay is recovered in our measurements for an OPD $\delta>70$ $\mu$m, \textit{i.e} when the diffusive regime is reached. This observation demonstrates that the measured Green's functions follow the temporal behavior predicted by diffusion theory, thus confirming the validity of our approach. 

In summary, this study demonstrates for the first time the optical measurement of time-dependent Green's functions with low-coherence interferometry. As a proof-of-concept, we have first been able to retrieve the time-resolved ballistic and multiple scattering contributions of the Green's function between individual scatterers. This approach has also been successfully applied to a strongly scattering medium. A whole set of time-dependent Green's functions can be measured between each point of the surface of a scattering sample. The experimental access to this Green's matrix is potentially important in many applications of wave physics in complex media whether it be for imaging \cite{shahjahan}, characterization \cite{aubry2}, focusing \cite{popoff2010measuring,choi,popoff2,choi2} or communication \cite{derode_2003,popoff2010image} purposes. 

The authors are grateful for funding provided by LABEX WIFI (Laboratory of Excellence within the French Program Investments for the Future, ANR-10-LABX-24 and ANR-10-IDEX-0001-02 PSL*). A.B. acknowledges financial support from the French ``Direction G�n�rale de l'Armement''(DGA). A. A. and G. L. would like to acknowledge funding from High Council for Scientific and Technological Cooperation between France and Israel under reference P2R Israel N$^o$ 29704SC.


\end{document}